\pdfoutput=1

\documentclass[twocolumn,showpacs,preprintnumbers,
prb,superscriptaddress]{revtex4-2}%
\usepackage{amssymb}
\usepackage{amsmath}
\usepackage{graphicx}
\usepackage{svg}
\usepackage{dcolumn}
\usepackage{bm}
\usepackage{amsfonts}%
\providecommand{\U}[1]{\protect\rule{.1in}{.1in}}

\begin{document}
\title{Ensemble Green's function theory for interacting electrons with degenerate ground states}
\author{E. Linn\' er$^1$ and F. Aryasetiawan}
\affiliation{Department of Physics, Division of Mathematical Physics, Lund University, Professorgatan 1, 223 62 Lund, Sweden}
\date{\today }

\begin{abstract}
An ensemble Green's function formalism, based on the von Neumann density matrix approach, to calculate one-electron excitation spectra of a many-electron system with degenerate ground states is proposed. A set of iterative equations for the ensemble Green's function and self-energy is derived and a simplest approximation corresponding to an ensemble \emph{GW} approximation is naturally obtained. The derivation is based on the Schwinger functional derivative technique and does not assume any adiabatic connection between a noninteracting and an interacting ground state.

\end{abstract}

\pacs{71.20.-b, 71.27.+a}
\maketitle

\section{INTRODUCTION}
A wide range of electronic systems found in nature have degenerate ground states. Prominent examples are open-shell atoms and molecules, vacancy defects in solids, two-dimensional electronic systems, quantum dots under magnetic field, and frustrated magnets. The degeneracy often gives rise to many fascinating phenomena not observed in systems with well-defined nondegenerate ground state. For example, the Landau degeneracy in a two-dimensional electronic system leads to fractional quantum Hall effect\cite{stormer1999} and the high degeneracy in frustrated magnets causes the system to fluctuate among the degenerate ground states even at temperature close to absolute zero, leading to emergent phenomena of fractional spin excitations and magnetic monopoles in spin ice \cite{balents2010,han2012}. Apart from the fundamental interest, systems with degenerate ground states may find useful applications in, for example, quantum computing.

For systems with nondegenerate ground state, there are already well-established methods developed over many years. Density functional theory (DFT) is a widely used method to calculate ground-state properties \cite{dreizler1990, parr1989} and Green's function method within many-body perturbation theory (MBPT), such as the \emph{GW} approximation (GWA), is routinely applied to study excited-state properties \cite{hedin65,aryasetiawan98,aulbur00,onida02}. The situation is completely different in the case of systems with degenerate ground states. Although DFT has been extended to the degenerate case, it has not been applied extensively \cite{heinonen1995}. There is even less work in developing methods for computing excited-state properties of systems with degenerate ground states.

An early attempt to extend the Green's function method to the degenerate case, without any concrete computational procedure provided, is by Layzer in 1962 \cite{layzer1962}. Later attempts of extending the method is the work of Cederbaum \emph{et al }in 1970's, in which they considered open-shell atoms and molecules \cite{cederbaum1974,cederbaum1977}. Several works applying the GWA to systems with degenerate ground states have appeared recently in the literature. Attaccalite \emph{et al.} \cite{attaccalite2011} and Ma \emph{et al.} \cite{ma2010} applied the GWA to defects in crystals but the problem with degeneracy associated with the open shell was not explicitly considered. Lischner and co-workers assumed a certain form for the self-energy and a careful choice of the starting mean field \cite{lischner2012}. An earlier work by Shirley and Martin avoided the degeneracy problem by special selection of the reference state \cite{shirley1993}. So far there is no general formulation based on the Green's function to treat systems with degenerate ground states. One of the main problems stems from MBPT that usually assumes an adiabatic connection between the true interacting ground state and a noninteracting ground state. This connection is no longer obvious for degenerate ground states. An alternative Green's function method for the degenerate case built upon the nonperturbative adiabatic approximation is by Brouder \emph{et al.} \cite{brouder2009}.

\section{THEORY}
In this paper, a Green's function theory based on the ensemble density matrix formalism in quantum mechanics pioneered by von Neumann in 1927 is proposed. An ensemble is characterized by the density matrix
\begin{equation}
\hat{D}=\sum_{i=1}^{M}w_{i}\left\vert \Phi_{i}\right\rangle \left\langle
\Phi_{i}\right\vert ,\ \ \sum_{i=1}^{M}w_{i}=1,\ 0\leq w_{i}\leq1 ,
\end{equation}
containing the information needed to calculate physical properties of the ensemble \cite{sakurai}.
$M$ is arbitrary and each weight $ w_{i}  $ determines the fraction of the ensemble in state $  \left\vert \Phi_{i}\right\rangle  $, with the states $\left\{  \left\vert \Phi_{i}\right\rangle \right\} $ not necessarily being orthogonal. The ensemble average of any operator $\hat{O}$ is given by $\text{Tr}(\hat{D}\hat{O})$. For example, for the density operator we find
\begin{equation}
\rho(r)=\text{Tr}[\hat{D}\hat{\rho}(r)]=\sum_{i=1}^{M}w_{i}\left\langle
\Phi_{i}|\hat{\rho}(r)|\Phi_{i}\right\rangle =\sum_{i=1}^{M}w_{i}\rho_{i}(r),
\end{equation}
with $r=(\mathbf{r},\sigma)$. This density is referred to as ensemble density.

Similarly, we define an ensemble Green's function as follows:
\begin{equation}
G(1,2)=\sum_{n=1}^M w_{n}G_{n}(1,2), \label{GEns}
\end{equation}
where a short-hand notation $1=(r_{1},t_{1})$ etc. is used and auxiliary Green's functions $G_n$ are
defined in the interaction picture according to
\begin{equation}
iG_{n}(1,2)=\frac{\left\langle \Psi_{n}|T[\hat{S}\hat{\psi}_{D}(1)\hat{\psi}_{D}^{\dag}(2)]|\Psi_{n}\right\rangle}{\left\langle \Psi_{n}|\hat{S}
|\Psi_{n}\right\rangle}, \label{Gint}
\end{equation}
where $T$ is the time-ordering operator and
\begin{equation}
\hat{S}=T\exp\left[-i\int_{-\infty}^{\infty}d\tau\int dr\hat{\rho}(r,\tau)\varphi(r,\tau)\right].
\end{equation}
$\left\{  \left\vert \Psi_{n}\right\rangle ,n=1,...M\right\}$ are arbitrary many-electron states, chosen later as the set of degenerate ground states, with corresponding fractions $w_n$ in which the system is prepared at some initial time point. The perturbing field $\varphi(r,t)$ is a virtual field that is used as a tool to derive the self-energy and it will be set to zero after taking functional derivatives of $G_n$. The choice of denominator in Eq. \eqref{Gint} is motivated later. The ensemble expectation value of any one-particle operator can be obtained from the ensemble Green's function as follows:
\begin{align}
\left\langle \hat{O}\right\rangle  &  =\sum_{n=1}^{M}w_{n}\left\langle
\Psi_{n}|\hat{O}|\Psi_{n}\right\rangle \nonumber\\
&  =-i\int dr\lim_{r^{\prime}\rightarrow r}O(r)G(rt,r^{\prime}t^{+}).
\end{align}

Each $G_n$ is of the same form as the nondegenerate Green's function, and thus the set of $G_n$ fulfills the set of equations of motion:
\begin{align}
&  \left(  i\frac{\partial}{\partial t_{1}}-h_0(1\mathbf{)}\right)
G_{n}(1,2)+i \int d3 v(1-3)G^{(2)}_n(1,2,3,3^+)\nonumber\\
&  =\delta(1-2), \label{eqnofmotionGn}
\end{align}
where the auxiliary two-particle Green's functions $G^{(2)}_n$ are defined in the interaction picture as:
\begin{equation}
G^{(2)}_{n}(1,2,3,4)\equiv -\frac{\left\langle \Psi_{n}|T[\hat{S}\hat{\psi}_{D}(1)\hat{\psi}_{D}^{\dag}(2)\hat{\psi}_{D}(3)\hat{\psi}_{D}^{\dag}(4)]|\Psi_{n}\right\rangle}{\left\langle \Psi_{n}|\hat{S}
|\Psi_{n}\right\rangle}. \label{G2int}
\end{equation}
Now utilizing the Schwinger functional derivative technique, the auxiliary two-particle Green's functions $G^{(2)}_n$ is related to the functional derivative of $G_n$:
\begin{equation}
    \frac{\delta G_n(1,2)}{\delta \varphi(3)} = i G_n(1,2)\rho_n(3) - G^{(2)}_n(1,2,3,3^+).
\end{equation}
A set of mass operators $M_n$, defined by
\begin{equation}
    i \int d3 v(1-3)G^{(2)}_n(1,2,3,3^+) \equiv -\int d3 M_n(1,3)G_n(3,2), \label{mass}
\end{equation}
are introduced, and will now be employed to rewrite Eq. \eqref{eqnofmotionGn}. By a further introduction of the ensemble Hartree potential $V^H$ and self-energies $\Sigma_n$:
\begin{equation}
V^{H}(1) \equiv \sum_{n=1}^{M}w_{n} V_{n}^{H}(1) \equiv \sum_{n=1}^{M}w_{n} \int d3v(1-3)\rho_{n}(3), \label{EVH}
\end{equation}
\begin{align}
\Sigma_n(1,2) & \equiv -i \int d3d4 v(1-3)G_{n}(1,4)\frac{\delta G^{-1}_n(4,2)}{\delta \varphi(3)} \nonumber \\
& + \delta(1-2)\left( V^H_n(1)-V^H(1)\right), \label{SE}
\end{align}
the mass operator can be written in the form:
\begin{equation}
M_n(1,2) = V^H(1)\delta(1-2) + \Sigma_n(1,2) . \label{mass2}
\end{equation}
By employing Eqs. \eqref{eqnofmotionGn} and \eqref{mass2}, the equation of motion can be reformulated as
\begin{align}
&  \left(  i\frac{\partial}{\partial t_{1}}-h(1\mathbf{)}\right)
G_{n}(1,2)- \int d3 \Sigma_n(1,3)G_n(3,2)\nonumber\\
&  =\delta(1-2), \label{eqnofmotionG}
\end{align}
with $h(1)=h_0(1)+V^H(1)+\varphi(1)$. From Eq. \eqref{eqnofmotionG}, we obtain the functional derivative of the inverse Green's functions $G^{-1}_n$,
\begin{equation}
    \frac{\delta G^{-1}_n(4,2)}{\delta \varphi(3)} = -\left( \delta(4-3) + \frac{\delta V^H(4)}{\delta \varphi(3)}\right)\delta(4-2) - \frac{\delta \Sigma_n(4,2)}{\delta \varphi(3)}.
\label{dGinverse}
\end{equation}
An important point is the choice of the ensemble Hartree potential $V^H$ instead of the individual $V_n^H$, thus modifying the structure of the self-energies. The motivation is that in the self-consistent mean-field Hamiltonian approach computations are in general only well defined for an ensemble mean-field Hamiltonian. The concept of degeneracy becomes ill defined when using separate $V^H_n$, as the label $n$ is arbitrary within a given set of degenerate states. Thus in the first iteration, we set $\delta\Sigma_n/\delta\varphi=0$ in Eq. \eqref{dGinverse} and $\delta G^{-1}_n/\delta\varphi$ is then determined only by the well-defined response of the ensemble Hartree potential $\delta V^H/\delta\varphi $, and is independent of the ill-defined $\delta V^H_n/\delta\varphi $.
We then obtain the self-energy
\begin{align}
\Sigma_n(1,2)  &  =iG_n(1,2)v(1-2)\nonumber \\
& + i\int d3 v(1-3)G_n(1,2)\frac{\delta V^H(2)}{\delta \varphi(3)}\nonumber\\
& +\delta(1-2)\left(V_n^H(1)-V^H(1)\right).\label{ESigma}
\end{align}

After having obtained $\Sigma_n$ in the first iteration, we form the ensemble self-energy 
$\Sigma=\sum_n w_n \Sigma_n$, which we use in the equation of motion of the Green's function as follows:
\begin{align}
&  \left(  i\frac{\partial}{\partial t_{1}}-h(1\mathbf{)}\right)
G_{n}(1,2)- \int d3 \Sigma(1,3)G_n(3,2)\nonumber\\
& -\int d3 \Delta\Sigma_n(1,3)G_n(3,2)
 =\delta(1-2), \label{eqnofmotionG1}
\end{align}
with $\Delta\Sigma_n= \Sigma_n - \Sigma$. From the above equation we find
\begin{align}
    \frac{\delta G^{-1}_n(4,2)}{\delta \varphi(3)} 
    &= -\left( \delta(4-3) + \frac{\delta V^H(4)}{\delta \varphi(3)}\right)\delta(4-2) - \frac{\delta \Sigma(4,2)}{\delta \varphi(3)}\nonumber \\
    &- \frac{\delta \Delta\Sigma_n(4,2)}{\delta \varphi(3)}.
\end{align}
We then use the above $\delta G^{-1}_n/\delta \varphi$ 
in Eq. (\ref{SE}), with
$\delta \Delta\Sigma_n /\delta \varphi$ set to zero since this quantity is not known at this iteration and consistent with the fact that it depends on $\delta V^H_n / \delta \varphi$, which in contrast to $\delta V^H / \delta \varphi$ is ill defined. The procedure can be continued to obtain higher order vertex corrections.
Unlike the original Hedin's equations which form a self-consistent loop, the corresponding equations in the degenerate case should be regarded at each iteration as a new perturbation expansion based on the previous degenerate Green's functions. The degeneracy may indeed be lifted in general yielding a new set of degenerate Green's functions. It is noteworthy that 
$\Delta\Sigma_n$ is zero in the nondegenerate case.

We now apply the above formalism to a system with degenerate ground states. The states
$\left\{  \left\vert \Psi_{n}\right\rangle ,n=1,...M\right\}$ are chosen to be degenerate ground states with energy $E_0$. The corresponding weights $\{w_n\}$ are set equal and given by $1/M$. For this choice, the ensemble Green's function contains the information for the ensemble average ground state energy, by the Galitskii-Migdal formula, and the one-particle excitation spectra. It is noteworthy that since the denominator in Eq. (\ref{Gint}) is equal to unity, the definition of the ensemble Green's function for degenerate ground states is invariant under a unitary rotation within the degenerate subspace.

A key quantity in calculating the self-energy is the density response function, which in turns determines the screened interaction. Thus, within the ensemble analog of the well-established GWA, which corresponds to setting $\delta \Sigma_n/\delta \varphi = 0$, leading to Eq. (\ref{ESigma}), the linear density response function is required in order to compute $\delta V^H/\delta \varphi $:
\begin{equation}
\frac{\delta V^H(1)}{\delta \varphi(2)}
=\int d3 v(1-3)R(3,2)
\end{equation}
with the linear density response function
\begin{equation}
R(1,2) \equiv \frac{1}{M}\sum_{n=1}^{M}R_n(1,2) \equiv \frac{1}{M}\sum_{n=1}^{M} \frac{\delta \rho_n(1)}{\delta \varphi(2)}.
\end{equation}
Introducing the basis $b_{\alpha}({\bf r}) = \phi^{*}_{i}({\bf r})\phi_{j}({\bf r})$, where $\phi_i$ is the orbital associated with $c_i$, the spectral representation of $R$ reads
\begin{equation}
R({\bf r,r'};\omega) = \sum_{\alpha \beta} b_{\alpha}({\bf r})R^{\alpha\beta}(\omega) b_{\beta}({\bf r}'), \label{Rmatrix}
\end{equation}
\begin{align}
R^{\alpha\beta}(\omega)  & = \frac{1}{M}\sum_{n=1}^{M}\sum_{m\neq n} \left[\frac{\rho^\alpha_{nm} \rho^\beta_{mn}}{\omega - E_m + E_0 + i\delta} \right. \nonumber \\
& \left.- \frac{\rho^\beta_{nm} \rho^\alpha_{mn}}{\omega + E_m - E_0 - i\delta} \right], \label{response}
\end{align}
where $\vert\Psi_m\rangle$ is an eigenstate of the Hamiltonian with eigenvalue $E_m$, where $\rho^\alpha_{nm} = \langle \Psi_n \vert \hat c_{i}^{\dagger} \hat c_{j} \vert \Psi_m \rangle$, and where $\alpha,\beta$ are the collective indices of $(i,j)$. In the above expression, the density operator does not couple a degenerate ground state labeled by $n$ to itself, due to the choice in \eqref{Gint}: When calculating the response function as a functional derivative of the Green's function or the density, the presence of the denominator cancels the term corresponding to the coupling of the density operator to the same state.
 
For systems with degenerate ground states, $m$ can specify other degenerate ground states, and the terms appearing in the degenerate subspace can thus diverge for $\omega \rightarrow 0$, which is reminiscent of the problem with the standard perturbation theory when applied blindly to a degenerate case.
A diagonalization procedure is proposed to eliminate this divergence. Diagonalizing the nonzero matrices $\rho^\alpha$, with $\rho_{nn}^{\alpha} = 0$ for all $n$ (since the density operator does not couple to the same state), in the subspace of the degenerate ground states for each $\alpha$ one obtains a new basis set of degenerate ground states which diagonalize $\hat\rho^\alpha=\hat c_{i}^{\dagger} \hat c_{j}$. The diverging terms vanish in this new basis set since $\rho^\alpha_{mn}=0$ for $m\neq n$. For the special case of all $\{w_n\}$ equal to $1/M$, the remaining nonvanishing terms are independent of the choice of the degenerate ground state basis and the response function can thus be rewritten as
\begin{align}
R({\bf r,r'};\omega) & =  \frac{1}{M}\sum_{n=1}^{M}\sum_{m}^{\text{exci}}
\left[\frac{\langle \Psi_n \vert \hat{\rho}({\bf r}) \vert \Psi_m \rangle \langle \Psi_m \vert \hat{\rho}({\bf r'}) \vert  \Psi_n \rangle}{\omega - E_m + E_0 + i\delta}
\right. 
\nonumber \\
& \left.- \frac{\langle  \Psi_n \vert \hat{\rho}({\bf r'}) \vert \Psi_m \rangle \langle \Psi_m \vert \hat{\rho}({\bf r}) \vert  \Psi_n \rangle}{\omega + E_m - E_0 - i\delta} \right],
\label{Rab}
\end{align}
where the sum over $m$ is now strictly over excited states, such that no divergence occurs when $\omega=0$. We note that in practice the diagonalization procedure is actually not required for uniform weights since the denominators in Eq. \eqref{Rab} do not depend on $n$. This shows that for the ensemble response function with uniform weights, the standard formula can be used with any chosen set of degenerate ground states except that transitions among these degenerate ground states are removed. In the general case the weights corresponding to the new basis set are modified by the diagonalization procedure.

Contained within the time-ordered response function is the physical retarded response function. By the Kubo formula \cite{fetter}, the retarded linear ensemble density response function can be constructed as:
\begin{align}
iR^r(1,2) & = \frac{1}{M}\sum_{n=1}^{M} \langle \Psi_n \vert [\Delta \hat{\rho}_n(1),\Delta\hat{\rho}_n(2)] \vert \Psi_n \rangle \theta(t_1-t_2),\label{retres2}
\end{align}
where $\Delta \hat{\rho}_n(1) = \hat{\rho}(1)-\rho_n(1)$. An equivalent and standard form in literature of retarded response is obtained by exchanging $\Delta \hat{\rho}_n(1)$ with $\Delta \hat{\rho}(1) = \hat{\rho}(1)-\rho(1)$. In the spectral representation, peaks at $\omega=0$ originating from the degenerate subspace do not appear in the retarded response function. A time-ordered response function is defined to satisfy the relations:
\begin{align}
\text{Re}R({\bf r,r'};\omega) & = \text{Re}R^r({\bf r,r'};\omega), \label{rettimrelation1} \\ \text{Im}R({\bf r,r'};\omega)\text{sgn}(\omega) & = \text{Im}R^r({\bf r,r'};\omega).\label{rettimrelation2}
\end{align}
Only the time-ordered response function defined in Eq. \eqref{retres2} satisfies relations \eqref{rettimrelation1} and \eqref{rettimrelation2}. The proposed form in Eq. \eqref{response} is based on the form \eqref{retres2}, where the diagonalization procedure can be employed.

A widely used approximation to compute the response function is the random-phase approximation (RPA), on which the GWA is based. As input, the noninteracting response function corresponding to some mean-field Hamiltonian is needed. If the mean-field ground state is degenerate, the same diagonalization procedure as described above can be employed.

\section{APPLICATION TO MODEL SYSTEMS AS PROOF OF CONCEPT}
As a proof of concept and an illustration on how the formalism works in practice, we consider a hydrogen like system, occupied by six electrons, and a two-dimensional harmonic oscillator, occupied by four electrons. In the H-like system the $1s$, $2s$, $2p$, $3s$ orbitals are considered, with the interaction between the electrons given by $v({\bf r-r'})=1/\vert {\bf r-r'}\vert$. The noninteracting ground state is nine fold degenerate, with the $1s$, $2s$ orbitals filled and two electrons occupying the $2p$ orbital, while the interacting ground states is nondegenerate. In the 2D harmonic oscillator, only the six lowest energy orbitals are considered, with the electron-electron interaction given by a point interaction $v({\bf r-r'})=U\delta({\bf r-r'})$. The noninteracting ground state is four fold degenerate, while the interacting ground state is nondegenerate. For both systems, the noninteracting problem is solved with a mean-field ensemble Hartree potential. A comparison of a one-shot ensemble $G^0W^0$ approach to the exact solutions as well as a one-shot nonensemble $G^0W^0$ approach is made. In the nonensemble approach the degeneracy is neglected by computing new sets of energies for each of the separate Hartree potentials of the noninteracting ground states, with the noninteracting system chosen to correspond to a nondegenerate noninteracting ground state with the lowest energy. The ensemble and nonensemble Green's function and self-energy are computed within the GWA.

We first compute the noninteracting Green's functions $G_n^0$ and noninteracting response function or the polarization $P^0$. Once the polarization $P^0$ is obtained the rest of the computation follows a routine procedure of first calculating the screened interaction $W=v+vP^0 W$ and then the self-energies given by 
\begin{align}
\Sigma_{n}({\bf r, r'};\omega) & = i \int \frac{d\omega'}{2\pi}G_{n}({\bf r, r'};\omega + \omega') W({\bf r', r};\omega')\nonumber \\
& +\delta({\bf r-r'})
\left(V_n^H({\bf r})-V^H({\bf r})\right),
\end{align}
which can be computed with a similar procedure as in the nondegenerate case. The ensemble Green's function $G$ can be computed from the auxiliary Green's functions $G_n$, obtained from the set of Dyson's equations:
\begin{align}
G_n(1,2) & = G_n^0(1,2) + \int d3d4 G_n^0(1,3)\Sigma_n(3,4)G_n(4,2).
\end{align}
In addition, we compute the spectral forms of the ensemble response function and ensemble Green's function $S$ and $A$, respectively. Special care is required to include the occupied and unoccupied peaks with the correct sign in the computation of the ensemble spectral function $A$.

As an illustration, consider, for example, the 2D harmonic case with four electrons, two with spin up and two with spin down. Let $\phi_0$ be the lowest orbital with energy $\epsilon_0$ and $\phi_1$, $\phi_2$, be the first excited orbitals with degenerate energies $\epsilon_1=\epsilon_2$. Consider one of the degenerate configurations in which $\phi_0$ is occupied by a spin-up and a spin-down electron and
$\phi_1$ is occupied by a spin-up electron whereas $\phi_2$ is occupied by a spin-down electron. The spin-up and -down Green's functions corresponding to this configuration are then, respectively,
\begin{align}
    G_\uparrow (r_1,r_2;\omega) & = \frac{\phi_0(r_1)\phi^*_0(r_2)}{\omega -\epsilon_0 - i\delta} 
    + \frac{\phi_1(r_1)\phi^*_1(r_2)}{\omega -\epsilon_1 - i\delta} 
    \nonumber \\
    &+ \frac{\phi_2(r_1)\phi^*_2(r_2)}{\omega -\epsilon_2 + i\delta} + ...
\end{align}
\begin{align}
    G_\downarrow (r_1,r_2;\omega) & = \frac{\phi_0(r_1)\phi^*_0(r_2)}{\omega -\epsilon_0 - i\delta} 
    + \frac{\phi_2(r_1)\phi^*_2(r_2)}{\omega -\epsilon_2 - i\delta} 
    \nonumber \\
    &+ \frac{\phi_1(r_1)\phi^*_1(r_2)}{\omega -\epsilon_1 + i\delta} + ...
\end{align}
It is important that the spectral function, which is proportional to the imaginary part of the Green's function, is computed separately for each spin channel. If we instead sum over the up and down Green's functions and compute the imaginary part of this sum, terms such as
\begin{equation}
    \frac{\phi_1(r_1)\phi^*_1(r_2)}{\omega -\epsilon_1 - i\delta} + \frac{\phi_1(r_1)\phi_1^*(r_2)}{\omega -\epsilon_1 + i\delta}
\end{equation}
will remove the peaks corresponding to the occupied and unoccupied spectra of the up and down Green's function.
Therefore, we propose to compute the separate spectral functions $A_n$ for the corresponding spin-polarized $G_n$, with the poles clearly separated, and then computing the ensemble $A$ as the weighted sum over $A_n$.

\begin{figure}[h!]
\centering
\includegraphics[scale=0.45]{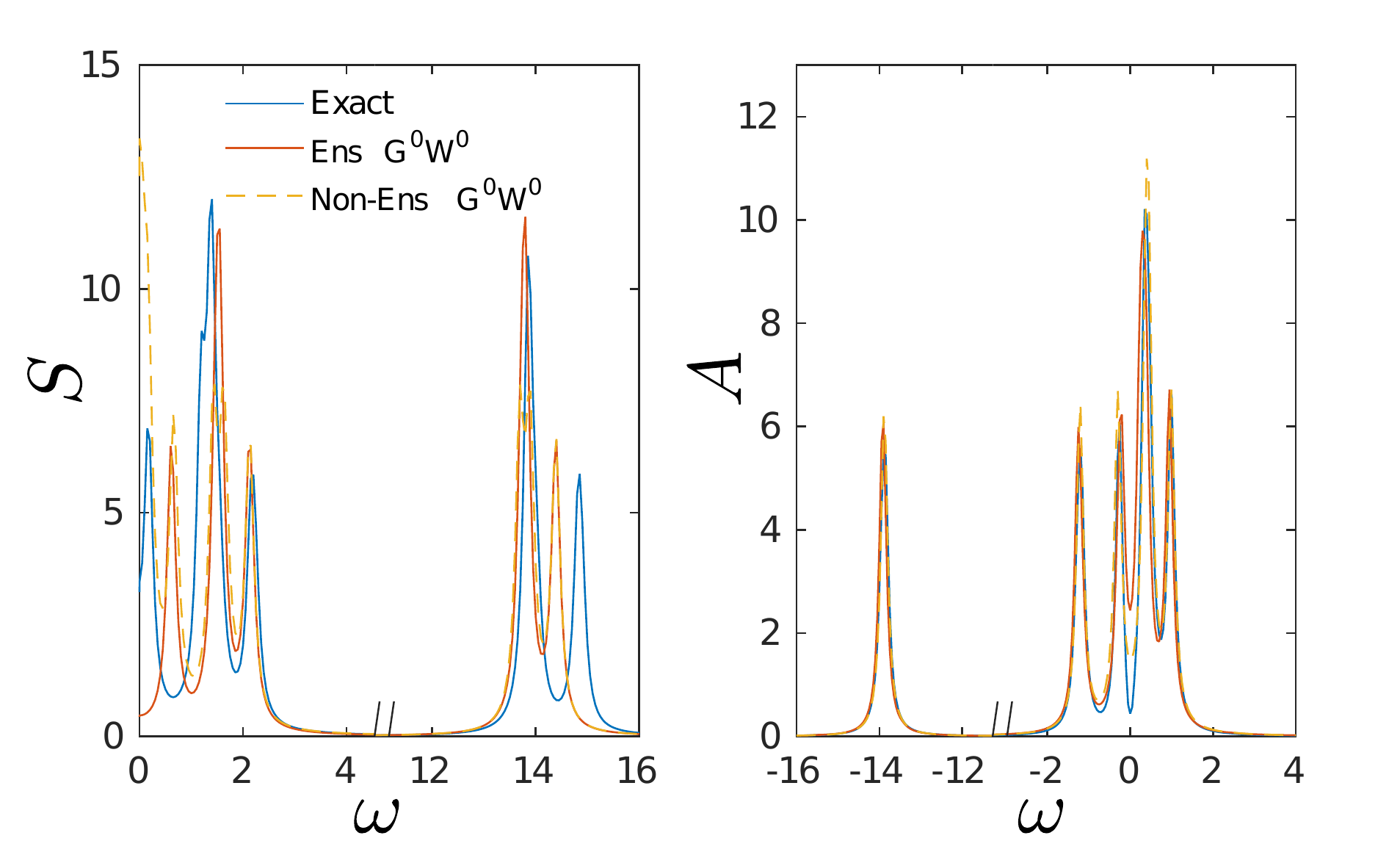}
\caption{The trace of the spectral response function $S$ (left figure) and spectral function $A$ (right figure) plotted against energy $\omega$ in the exact and ensemble and nonensemble $G^0W^0$ cases for the H-like system. Breaks in the x axes are employed.}\label{Figure1}
\end{figure}

In the H-like model the nuclear charge $Z=6$ and $Z=3$ is used for the initial one-electron energies and orbitals, respectively, as convergence issues appear in the noninteracting mean-field solution when the orbitals of the $Z=6$ system are used. The trace of $S$ and $A$ are plotted against $\omega$ in the three cases in Fig. \ref{Figure1}. In the ensemble case, the main exact peak structure of $S$ is well captured, except for an absence of the low energy peak structure which the nonensemble approach can partially capture. An excellent agreement of the peak structure and positions of $A$ between the exact and ensemble $G^0W^0$ approach $A$ is observed.

The low $\omega$ peak structure in $S$ corresponds to transitions originating from the degenerate noninteracting ground state subspace, which vanishes in the diagonalization procedure, and which may appear in the degeneracy breaking going from the noninteracting to the interacting system. The degeneracy breaking is first included in the self-energy, and thus the peaks are absent in the one-shot approach. A self-consistent approach is expected to be able to capture the absent peaks. After the first iteration the  self-energies $\Sigma_n$ will likely lower the starting symmetry, splitting the degenerate $2p$ states, which in the next iteration will yield the low-energy peaks. If we restrict ourselves to the one-shot approach, a mean-field Hamiltonian capturing the energy structure of the system better than the ensemble Hartree approach would be required to capture the low-energy peaks.

\begin{figure}[h!]
\centering
\includegraphics[scale=0.45]{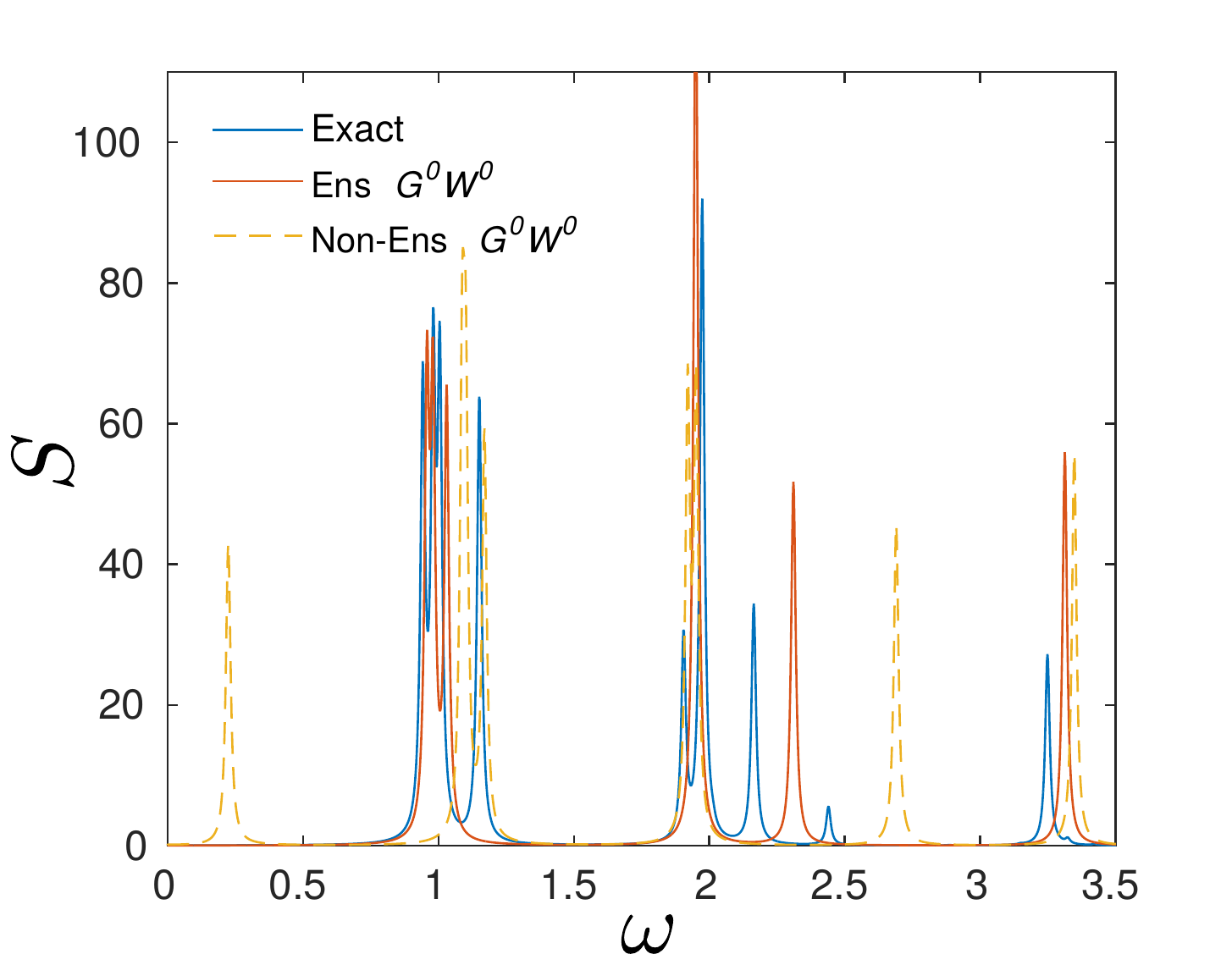}\\
\includegraphics[scale=0.45]{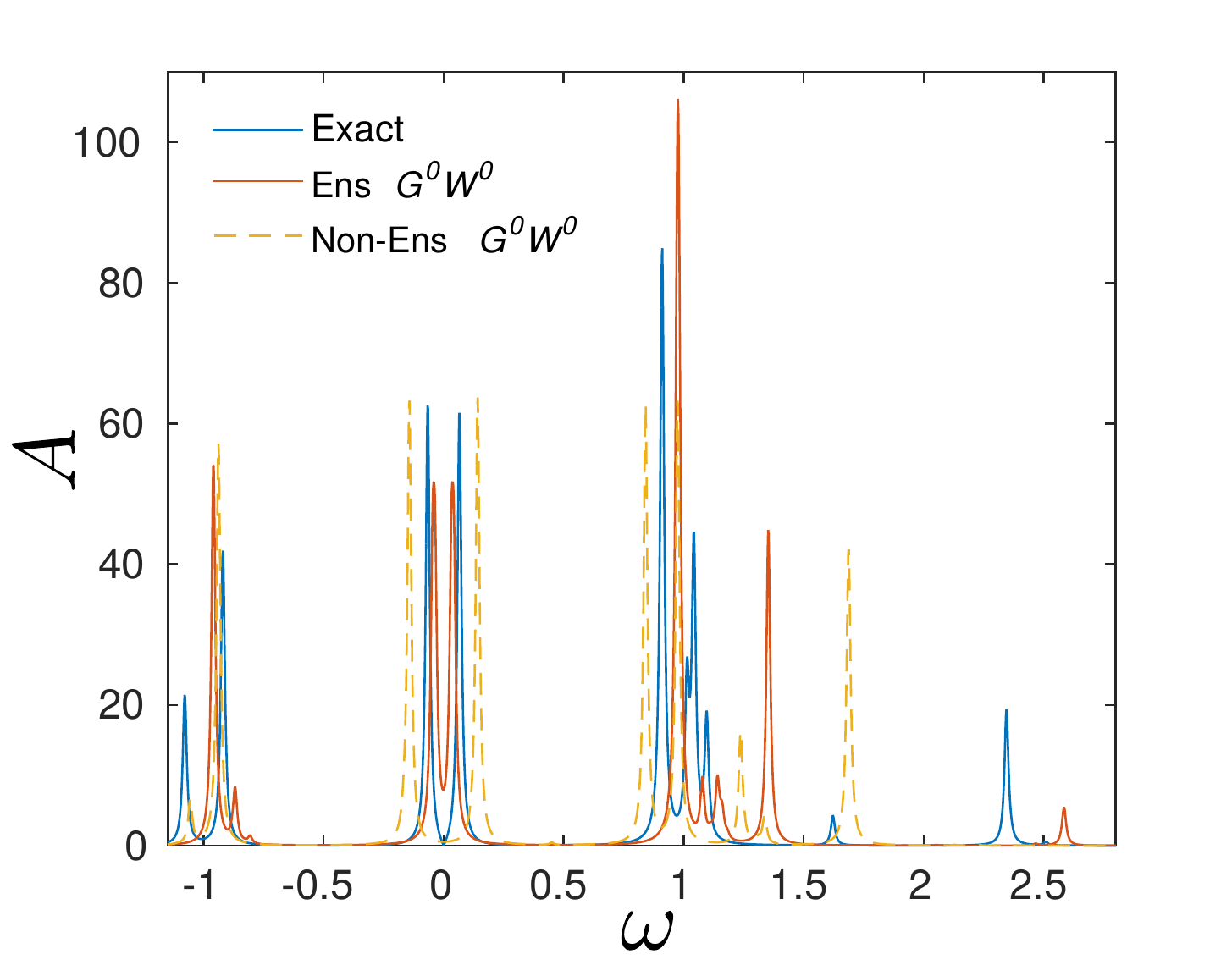}
\caption{The trace of the spectral response function $S$ (upper figure) and spectral function $A$ (lower figure) plotted against energy $\omega$ in the exact and ensemble and nonensemble $G^0W^0$ cases for the 2D harmonic oscillator. Small satellite features are not included in the plot.}\label{Figure2}
\end{figure}

The trace of $S$ and $A$ are plotted against $\omega$ for the 2D harmonic oscillator, with $U=1$, in the three cases in Fig. \ref{Figure2}. The ensemble peak structure of $S$ is in reasonable agreement with the exact one and in better agreement than the nonensemble approach. The nonensemble approach incorrectly predicts a peak at low $\omega$, while no low $\omega$ peak is present in the ensemble approach. A good agreement between the exact and ensemble $G^0W^0$ approach for the main peak structure of $A$ is observed, however, some detailed peak structure is captured better by the nonensemble approach, for example in the vicinity of $\omega=-1$.

\section{SELF-CONSISTENCY AND FINITE-TEMPERATURE TREATMENT}
An iterative self-consistent computational scheme for $G_n$ can be constructed. The polarization can be computed from the set of $G_n$ by the following ensemble analog of one of the Hedin's equations within GWA:
\begin{equation}
    P(1,2) = -\frac{i}{M}\sum_{n=1}^M G_n(1,2)G_n(2,1^+).
\end{equation}
The diagonalization procedure is employed for the computation of the polarization in each iteration. A conceptual issue is the degeneracy breaking in an iteration. We propose employing the Galitskii-Migdal formula on the auxiliary Green's functions to identify degeneracy, choosing the ones giving the lowest ground state energy for the following iteration. A slight mixing between the auxiliary Green's functions and the ensemble Green's function can be employed.

We propose extending the finite-temperature Green's function theory to include degenerate states by writing the Matsubara Green's function in the modified interaction picture as a weighted sum over auxiliary Matsubara Green's functions, with the weight given by the Boltzmann distribution. This choice leads to an ensemble real-time response function which satisfies the required properties of the time-ordered response function.

\section{CONCLUSION}
In summary, we have developed an ensemble Green's function formalism for treating many-electron systems with degenerate ground states in a well-defined way. A set of iterative equations, analogous to Hedin's equations for the nondegenerate case, is derived for the ensemble Green's function. An ensemble GWA is naturally obtained from the iterative equations. The formalism does not rely on an adiabatic connection between interacting and noninteracting ground states as commonly assumed in many-body perturbation approaches. Further application to realistic systems with degenerate ground states in the future would enlighten the strengths and weaknesses of the formalism. Most considerations were applicable for an arbitrary set of states $\vert \Psi_n \rangle$ and weights $w_n$, and studying other choices of ensembles capturing nonequilibrium aspects would be of interest.
\begin{acknowledgments}
We gratefully acknowledged financial support from the Swedish Research Council (VR).
\end{acknowledgments}

\end{document}